\documentclass[preprint,12pt,3p, numbers, authoryear]{elsarticle}
\usepackage{amssymb}
\usepackage{graphicx}
\usepackage{bm}
\usepackage{natbib}
\usepackage{multirow}
\usepackage{graphicx}
\usepackage{subcaption}
\usepackage{mwe}
\usepackage{xcolor}
\usepackage{lineno}
\usepackage{hyperref}

\journal{Neuron}

\begin{document}

\begin{frontmatter}

\title{Challenges for machine learning in clinical translation of big data imaging studies}

\author[label1]{Nicola K. Dinsdale\corref{cor1}}

\address[label1]{Wellcome Centre for Integrative Neuroimaging, FMRIB, Nuffield Department of Clinical Neurosciences, University of Oxford, UK}
\address[label2]{Institute of Biomedical Engineering, Department of Engineering Science, University of Oxford, UK}
\address[label3]{Australian Institute for Machine Learning (AIML), School of Computer Science, University of Adelaide, Adelaide, Australia}
\address[label4]{South Australian Health and Medical Research Institute (SAHMRI), North Terrace, Adelaide, Australia}

\author[label2]{Emma Bluemke}

\author[label1]{Vaanathi Sundaresan}

\author[label1,label3,label4]{Mark Jenkinson}

\author[label1]{Stephen M Smith}

\author[label2]{Ana I. L. Namburete}

\cortext[cor1]{Corresponding Author - nicola.dinsdale@dtc.ox.ac.uk}

\begin{abstract}
The combination of deep learning image analysis methods and large-scale imaging datasets offers many opportunities to imaging neuroscience and epidemiology. However, despite the success of deep learning when applied to many neuroimaging tasks, there remain barriers to the clinical translation of large-scale datasets and processing tools. Here, we explore the main challenges and the approaches that have been explored to overcome them. We focus on issues relating to data availability, interpretability, evaluation and logistical challenges, and discuss the challenges we believe are still to be overcome to enable the full success of big data deep learning approaches to be experienced outside of the research field.
\end{abstract}


\end{frontmatter}

%\linenumbers

\section{Introduction}
Across neuroimaging, the majority of datasets have been limited to small-scale collections, typically focusing on a specific research question or clinical population of interest. Recently, however, large scale `big data' collections have begun to be collated, many of which are openly available to researchers, meaning that, if the acquisition protocol, demographic and non-imaging information of the data meets the requirements of the study, novel research can be completed without having to acquire new scans. The sharing of these large-scale datasets has had many benefits: not only do they enable the exploration of new research questions, they also enable reproducibility and quicker method prototyping. 

Existing large-scale datasets have been curated to explore different research questions with varying numbers of subjects and imaging sites across studies. For instance, if the research question were about lifespan and ageing, datasets to consider would include: IXI\footnote{ \url{https://brain-development.org/ixi-dataset}} ($n=581$), NKI-RS \citep{Nooner2012} ($n>1000$), UK Biobank \citep{Biobank} ($n=100,000$), CamCAN \citep{Taylor2017} ($n=700$) and Lifespan HCP in Ageing  \citep{Bookheimer2018} ($n>1200$). Similarly, if one was interested in early development, available datasets include: dHCP \citep{Hughes2017} ($n>1000$) and ABCD \citep{Marek2019} ($n>10,000$), or for research on young adults, one could consider: HCP Young Adult \citep{VanEssen2013} (n=1200), GSP \citep{Holmes2015} ($n=1070$) and CoRR \citep{Zuo2015} ($n=1629$). Datasets also exist that explore specific clinical groups, such as Alzheimer's (OASIS 3 \citep{Marcus2007} ($n=1098$) and ADNI \citep{ADNI} ($n > 3500$)), Autism (ABIDE \citep{ABIDE} ($n=1112$)), and schizophrenia and bipolar disorder (CANDI \citep{Frazier2008} ($n=103$)). These datasets allow the exploration of questions that would not be possible with traditional small-scale studies where, for instance, the statistical power would not be sufficient to find significant results. For instance, if we are interested in exploring individual differences from the average trajectory rather than differences within-subject or group, we are likely to require a larger sample size afforded by the large-scale datasets \citep{Madan2021}. Large-scale studies have also enabled the characterisation of potential subtypes within patient samples -- for example, \citep{Young2018} demonstrated heterogeneity and subtypes in atrophy patterns due to Alzheimer's disease using data from ADNI. 

UK Biobank \citep{Biobank} is the largest of these studies, with a goal of collecting brain imaging data from 100,000 volunteers, including 6 MRI modalities, to study structure, function, and connectivity. UK Biobank also contains large quantities of lifestyle, genetic and health measures, which allow researchers to create models of population ageing and model how genetic and environmental factors interplay with this process. For instance, the atrophy of the hippocampus is a well-validated biomarker for Alzheimer's disease, and so, using the UK Biobank, a nomogram of hippocampal volume with normal ageing has been created \citep{Nobis2019}, illustrating the progression with age and percentiles of expected volume across the population. 

For population-based models to have maximal impact they need to be \textit{translatable} -- that is, they need to have genuine clinical impact beyond the research field. Having created a model of healthy ageing, when a patient arrives in the clinic we wish to be able compare their MRI scan to normative distributions. This would then allow us to identify whether the patient was ageing differently to the population average, and potentially identify the need for intervention. Unfortunately, comparing a scan to the normative model is not as simple as merely acquiring a scan and making that comparison. First, research data differs from clinical data in terms of quality, acquisition and purpose and, further, MRI data acquired on different scanners or with different protocols may differ in characteristics to such an extent that comparisons with the model are no longer valid. These `batch effects' also cause a harmonisation problem within many large studies, where data collected on different scanners within the study contain bias introduced by the effects of the particular scanner hardware and acquisition protocol on the image characteristics. Second, the populations may differ so much that the patient falls outside of the population modelled, and it cannot be  known if the models will extrapolate. For instance, as UK Biobank spans subjects from 45-85 years of age, any model developed on this data is unlikely to extrapolate well to subjects much younger than that range. Third, logistical challenges make the deployment of the models difficult in a clinical setting, potentially to the point where it is not currently possible for them to be used outside of the research world, for instance the dependence on GPUs for processing which are unlikely to be available. Further, many research studies utilise functional and diffusion MRI which are often unavailable clinically due to the lack of expertise and equipment, and long acquisition times being impractical (e.g., to match the quality or time taken over research scans). Functional connectivity analyses, for instance, have become a prominent approach for examining individual differences, driven by the availability of high quality data from the large-scale studies. Ultimately, if this data cannot be acquired clinically, the utility of any model is limited. Solutions to these problems, however, are beginning to be developed and deep learning offers potential opportunities.

Due to the growth in size of the datasets, deep learning models are now an option for neuroimaging analysis, enabling us to explore new questions in a data-driven manner. Applications of deep learning techniques to neuroimaging data have been explored in a research setting, with increasing numbers of novel methods proposed year on year. Powered by their ability to learn complex, non-linear relationships and patterns from data, deep learning methods have been applied to a wide range of applications and have found success in previously unsolved problems. However, challenges for applying deep learning models to the clinical domain remain. These challenges limit the impact that big datasets such as the UK Biobank are currently able to have on patient care, and work must be undertaken to allow models to extend beyond the research domain. Recent developments in deep learning have begun to tackle the problems faced, but further developments are needed. Here, we will discuss the challenges being faced and current approaches being developed to mitigate them, covering challenges around data availability, interpretability, model evaluation and logistical challenges including data privacy. We will also identify and explore the barriers we believe still need to be overcome. 

\begin{figure}
        \centering
	\includegraphics[width=\textwidth]{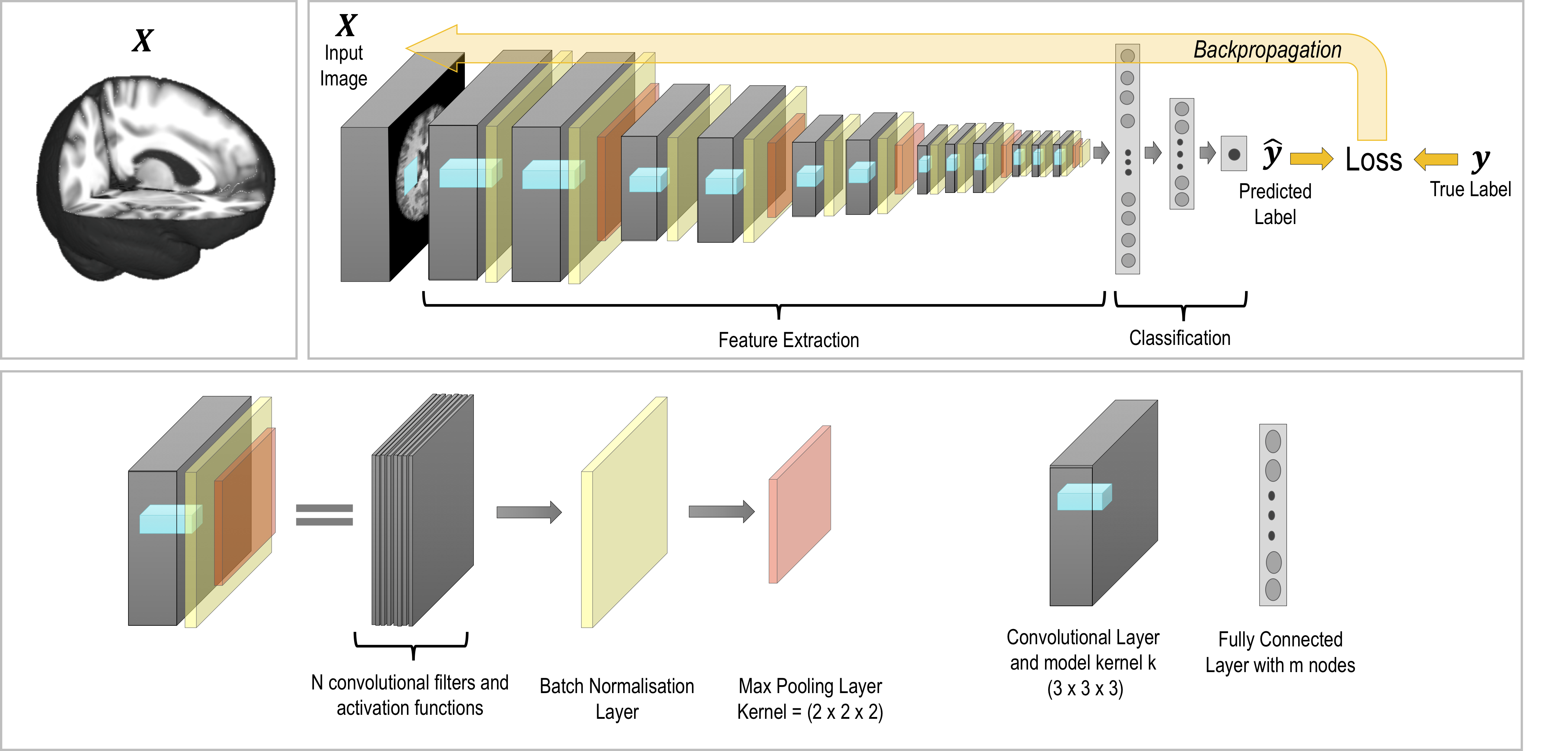}
	\caption{An example network architecture for a convolutional neural network (CNN) for a classification or regression task. }
	\label{fig:network}
\end{figure}

\subsection{Deep Learning Background}
To understand the challenges for clinical translatability of deep learning methods, we first require an overview of how these methods approach problems -- for a more detailed introduction see, for example, \citep{LeCun2015}. We will only consider convolutional neural networks (CNNs), which form the vast majority of deep learning methods currently applied in medical imaging, an example architecture of which is shown in Fig. \ref{fig:network}. The majority are \textit{supervised} approaches \citep{LeCun2015}, meaning that to explore the research question, we need access to a dataset of images, $\bm{X}$, and the set of known true labels, $\bm{y}$, for the task we wish to explore. This requires an understanding of the information that we expect to be encoded within the images and an understanding of which questions are of interest, defined as \textit{domain knowledge}. Examples for the $\bm{X}$ and $\bm{y}$ data could be a structural scan with an associated segmentation mask, or multiple modality data for $\bm{X}$, with the label being disease prognosis. 

Having curated and appropriately preprocessed the data (for instance, skull stripping, bias field correction, see e.g. \citep{Manjon2017}) the task is then to design a neural network architecture which we expect to be able to map from $\bm{X}$ to $\bm{y}$ through learning a highly non-linear mapping function $f(\bm{X}, \bm{y}; \bm{W})$, where $\bm{W}$ are the trainable weights of the neural network.  The choice of architecture is highly influenced by a variety of factors: the task being explored, the quantity of data available, and the computational power available, and is again highly influenced by domain knowledge. 

Nevertheless, most networks are formed through the same basic building blocks. The first are \textit{convolutional filters}, which learn features of interest from the data, encoding spatial relationships between pixels by learning the image features from small patches of the input data, determined by the filter's \textit{kernel size}. These layers, then, complete \textit{feature extraction}. They contain the weights and biases which need to be learned during the optimisation process, $\bm{W} = \{\bm{w}, b\}$, and stacks of these layers are placed at different spatial resolutions in order for a range of different features to be extracted at each level of abstraction, allowing us to create a rich understanding of the input data. During the forward pass of the \textit{back propagation} training procedure \citep{backpropagation}, each filter is convolved across the width and height of the input volume. This means that the network learns filters which activate when specific features are detected, and the exact nature of the features is learned through a network optimisation procedure that updates the filter weights, in order to find features that are useful contributors to the overall goal of predicting $\bm{y}$. 

\begin{figure}
        \centering
	\includegraphics[width=0.9\textwidth]{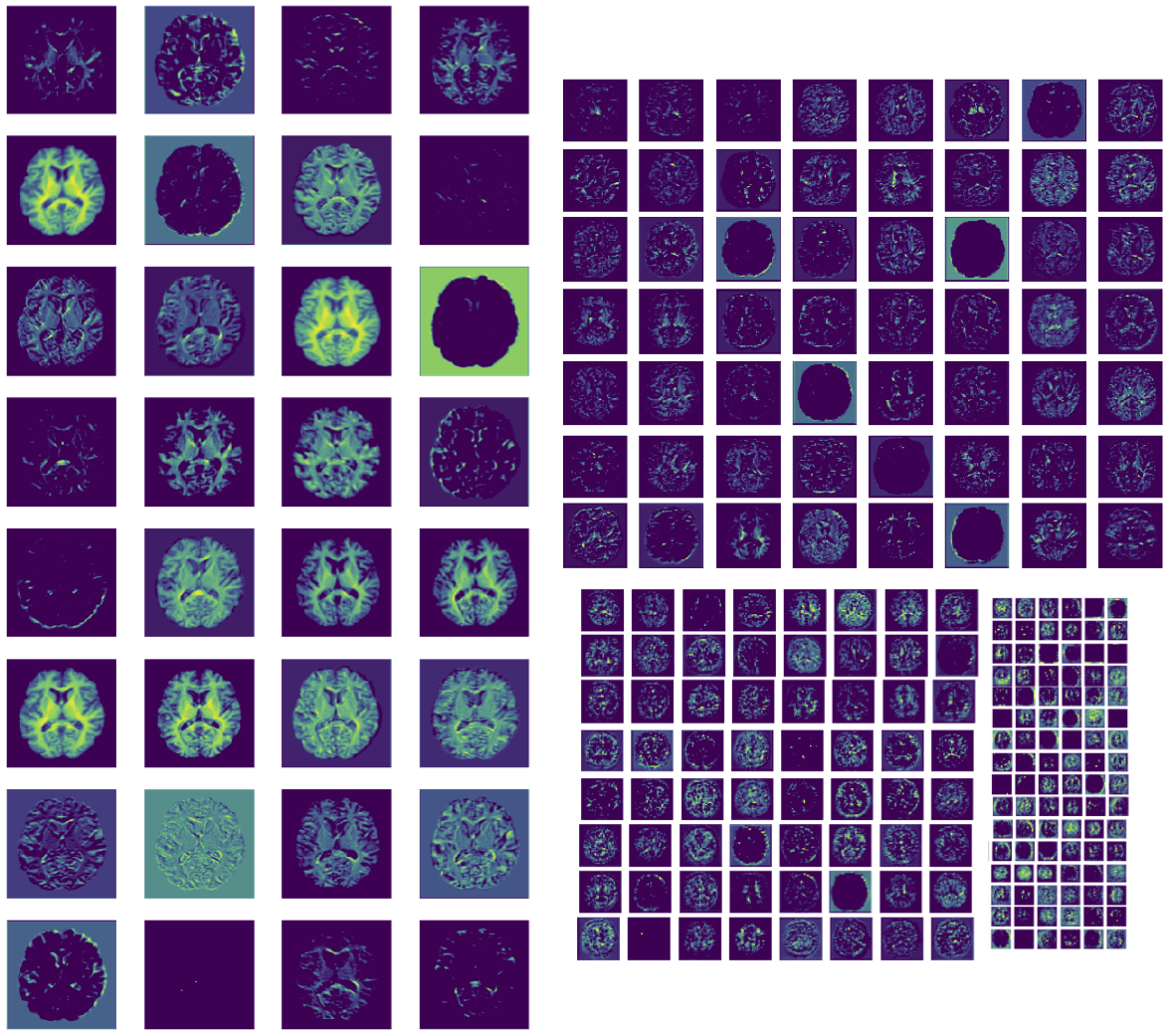}
	\caption{Example extracted features, $\bm{z}_l$, for brain age prediction (see \citep{Dinsdale2021}) at different network depths and spatial resolutions. It can be seen that several different features are extracted and some features are repeated, showing redundancy in filters.}
	\label{fig:features}
\end{figure}

The next components are the \textit{activation functions} which play a fundamental role in the model training. The activation functions apply a non-linear transformation to the data, after it has been weighted by the convolutional layers. This non-linearity provides a distinct edge to CNNs, allowing them to learn the complex non-linear relationships (or mapping) between the input and the output. Without the non-linear activation functions, CNNs would be rendered as only linear models, despite the flexibility to learn the weights.  Commonly used activation functions include rectified linear units (ReLU, e.g., zeroing negative values and keeping positive ones unchanged) \citep{ReLU} and sigmoid (e.g., squashing large values down to ceilings). Therefore, the features, $\bm{z}_l$, at layer depth $l$ are given by $\bm{z}_l = ReLU(\bm{z}_{(l-1)} * \bm{w}_l + b_l)$ and we can see that, due to the CNN's sequential data flow, the features at a given depth are a non-linear combination of the previous features and the weights and biases. Therefore, despite the sequential nature of CNNs, without the activation functions we would only be able to train linear models. Example extracted features, $\bm{z}_l$, after the activation function, can be seen in Fig. \ref{fig:features}.

The networks then learn features at different spatial resolutions through the inclusion of \textit{pooling blocks}, which reduce the input over a certain region to a single value, thus sub-sampling the data and condensing information. Pooling  condenses the intensity-based information, provides a basic invariance to rotations and translations, and has been demonstrated to improve the object detection capability of convolutional networks. Learning features at different resolutions allows the network to create a rich understanding of the input image, aiding the mapping between the input data, $\bm{X}$, and the output label, $\bm{y}$. 

The final key components of neural networks are \textit{fully connected layers}, which are essential to many classification or regression architectures. They are normally placed at the end of a network, after the convolutional layers have extracted features from the data, and then the fully connected layers learn how to classify these features.  The activation map from the final layer of the feature extraction  is reshaped into a long vector instead of a volume \textit{tensor}. In fully connected layers, all nodes in one layer are connected to all the nodes in the next, meaning that they are much more computationally powerful and expensive than convolutional layers.

By feeding the data through the network, we are able to create an output prediction. To make these predictions accurate, the weights of the network must be optimised using a process called \textit{back propagation} \citep{backpropagation}, and normalisation techniques such as \textit{batch normalisation} are added to minimise the impact of outliers and help result generalisation (i.e. performance on new or unseen datasets)\citep{Batchnorm2015}. To this end, we evaluate a \textit{loss} or \textit{cost function} which determines the error in the network prediction by comparing the prediction ($\bm{\hat{y}}$) and the true label ($\bm{y}$). The choice of the loss function is task-dependent, and plays a crucial role in the final network performance. The optimisation algorithm is often stochastic gradient decent or a similar algorithm, which updates the weights such that we reach a \textit{minimum} in the \textit{loss space}: that is, a model which minimises the loss. The aim is to find the \textit{global minimum} of the loss space, which is the place in the loss space which most minimises the loss function. However, the optimisation can get stuck in \textit{local minima}, which are spots of low loss due to the non-convex nature of the loss function, and so the \textit{learning rate} (the step size taken during the optimisation process) must be chosen to best help the model find the global minimum. 

\begin{figure}
        \centering
	\includegraphics[width=0.7\textwidth]{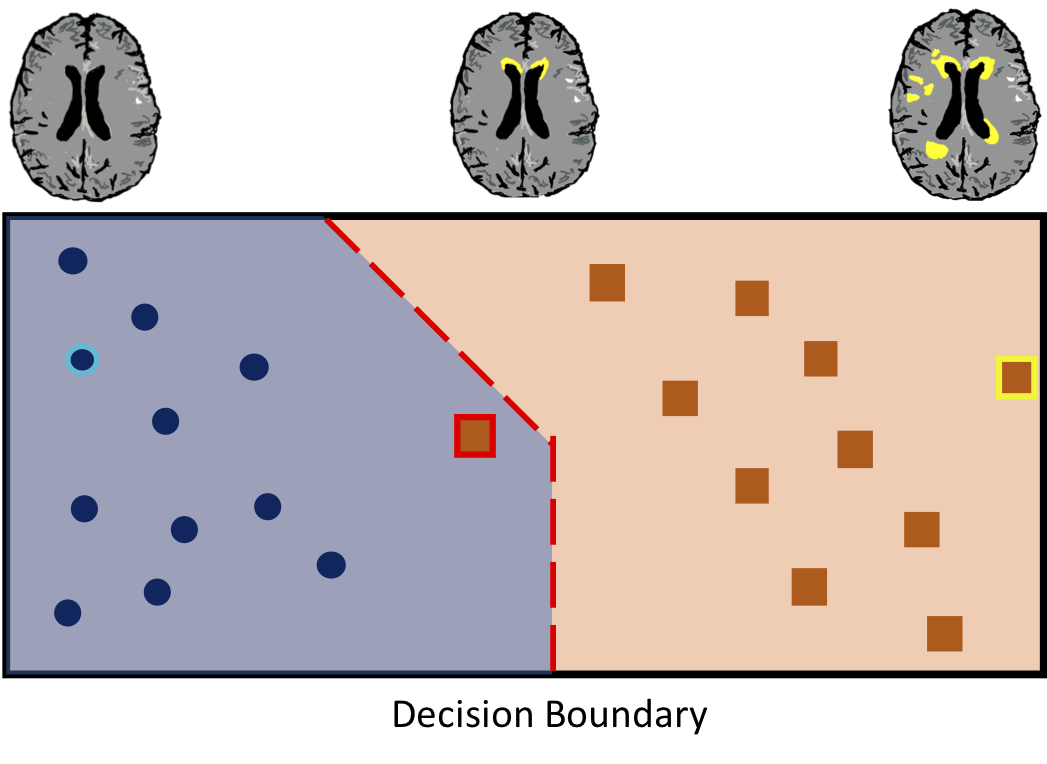}
	\caption{When Learning to classify samples (e.g., pixels or subjects - circles and squares), there is likely to be a continuum between the classes and the decision boundary chooses where one class begins and another ends. For subjects close to this decision boundary, it becomes critical as it determines the classification given. }
	\label{fig:decisionboundary}
\end{figure}

Therefore, we have an optimisation problem, the performance of which is highly dependent on two factors: first, the design decisions made about the network architecture and the loss function; second, the data available to train the network. This optimisation process determines the location of the \textit{decision boundary}, that is which locations in the feature space will be allocated to which class and so which data points are allocated to each class, as illustrated in Fig. \ref{fig:decisionboundary}.

In computer vision, where most of the deep learning techniques have been developed, very large datasets are available, with datasets commonly consisting of many millions of data points (e.g., 2D training images with labels), and being relatively easy to curate, for instance, by scraping the internet for images. In neuroimaging in general, data has to be labelled by a domain expert rather than mechanically produced. This is one of many differences between working in medical  imaging and the general computer vision field, and therefore, although many of the methods used  were developed in vision-related fields, there are challenges specific to working in the medical imaging or neuroimaging domain.

\section{Data Availability}

For clinical translatability of deep learning techniques, data availability is a major limitation. Despite the growth in the size of available datasets, the largest are still only of the order of tens of thousands of imaging subjects, with a thousand images being commonly regarded as a large dataset. For many specific tasks, datasets exist only in the order of hundreds of subjects, due to many factors such as the monetary and time cost of acquiring data, the difficulties in sharing and/or pooling data across sites, and the fact that, for some conditions, there are insufficient patients to create a dataset of any great size \citep{Morid2021}. For instance, the frequently explored Brain Tumour Segmentation (BraTS) dataset  \citep{BRATS} only has data from 369 subjects\footnote{Training Data for 2020 challenge} available for training, which is in stark contrast to the popular datasets from computer vision such as ImageNet \citep{imagenet} (1,281,167 training examples), CIFAR-10 \citep{Cifar10} (50,000 training examples) and MNIST \citep{mnist} (60,000 training examples) where many of the methods are being developed. By simply looking at dataset sizes, it is clear that we are likely to be underpowered for training neural networks. Highly parameterised, deep neural networks are very dependent on the amount of available training data \citep{Cho2015HowMD, He2020Datapoints}, with performance generally improving as the number of data points is increased, and they are far more affected by the amount of available training data than classical machine learning techniques, due to the need to learn the useful features as well as the decision boundary \citep{He2020Datapoints}.

\subsection{Maximising the impact of  available data}
There has, therefore, been an increasing focus on developing techniques to facilitate the  most effective use of the data available. A commonly used technique from computer vision is the use of large natural image datasets \citep{Raghu2019,Talo2018,Mehmood2021}, with ImageNet \citep{imagenet} being the most popular \citep{Cheplygina2019}, to \textit{pre-train} the network. Pre-training involves training the weights on a related task with more available data, such that the optimisation starts from an informed place, rather than a random initialisation. We can see why this might be useful by considering the information learned by the network at the different stages \citep{olah2018}: the early layers learn features such as edges and simple textures, largely resembling Gabor filters -- see Fig. \ref{fig:features} -- and are therefore very general and applicable across different images, regardless of the target tasks \citep{Yosinski2014}. The deeper convolutional layers then tend to learn features such as more complex textures and object  parts, and the final layers learn features which are far more task- and dataset-specific (e.g. fully connected layers learn discriminative features for a classification task). Therefore, we can take a network pre-trained on the large, canonical dataset, and either use the network to extract features which we then pass to a classifier, requiring only the classifier to be trained, or, more commonly, we can fine-tune (re-train) the deeper layers to the specific task. This requires less data, as not only are we starting the optimisation process from an informed point in the \textit{parameter space}, but also the very earliest layers can often be frozen (kept at their value and not updated during training), greatly reducing the number of weights in the model that need to be optimised. This process is referred to as \textit{transfer learning} and is a step frequently used to allow networks to be trained with lower amounts of training data. Transfer learning can be performed across data domain (dataset), task, or both, depending on the datasets available for pre-training. 

The standard practice is to use the very large datasets of natural images for the pre-training. However, natural images have very different characteristics from many medical images, and thus the features learned are not necessarily the most appropriate for the tasks we need to consider \citep{Raghu2019}. For instance, natural images are often stored as RGB images, whereas MR images are encoded as greyscale. Similarly, in medical images the location of structures could be informative, which is rarely true in natural images. Creating pre-trained networks for medical images has therefore been a focus, with Model Genesis \citep{ModelGenesis} creating a flexible architecture trained to complete multiple tasks, to create features which aim to generalise across medical imaging tasks. Similarly, some works pre-train on large datasets such as UK Biobank for tasks such as age or sex prediction, where obtaining labels is relatively trivial \citep{Lu2021} or on datasets for the same task with a dataset where more labels are available \citep{Kushibar2019,  Bashyam2020, Abrol2020}.  Here again, the aim is to learn features from the prediction task which are useful for the task we are interested in: that is, features that generalise across tasks. 	

Other studies utilise self-supervised approaches, such as \textit{contrastive representation learning}, where general features of a dataset are learned, without labels, by teaching the model which data points are similar or different. These then act as the starting point for model training on the smaller target dataset, rather than pre-training the model on a different dataset. An example approach is presented in \citep{SIMCLR}, where the data has been \textit{augmented} (small transformations applied to increase the size of the dataset, discussed below) and then the network is trained to encode the original image and the augmented image into the same location in the feature space  (i.e., they make the same output prediction), using a contrastive loss function \citep{Hadsell2006}, that learns features describing the similarity between images. Different self-supervised methods and contrastive loss approaches have been developed \citep{hanaff2020, Grill2020}, and have begun to be applied for medical imaging applications \citep{Zhang2020Contrastive, Chaitayna2020}, including for segmentation of MRI scans of the brain \citep{Zhuang2019, Chen2019}, increasing the performance on tasks where a small amount of training data is available. 

\subsection{Data Augmentation}

\begin{figure}
        \centering
	\includegraphics[width=0.75\textwidth]{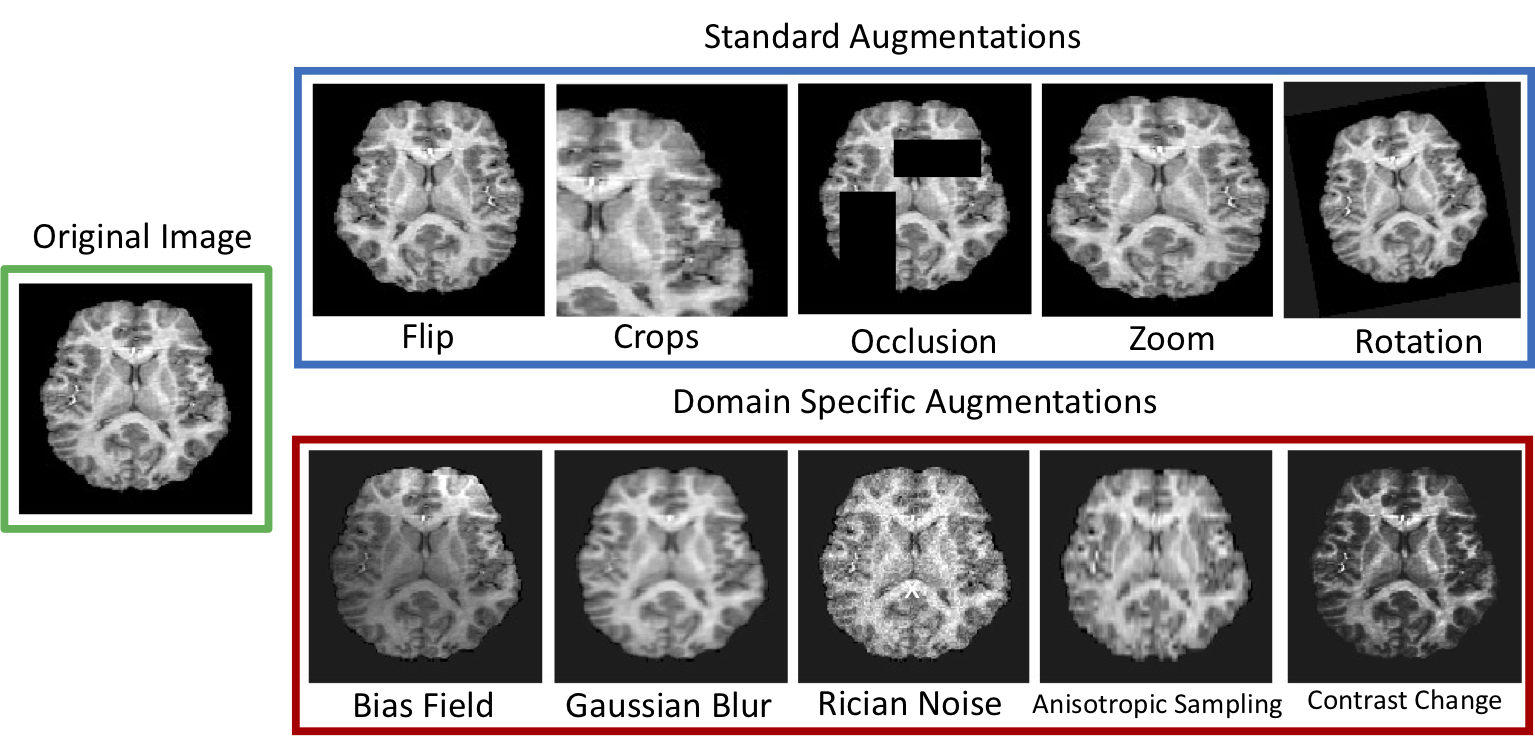}
	\caption{Example augmentations that might be applied to a MRI image. Standard augmentations, those that come directly from computer vision approaches, and domain specific augmentations for neuroimaging which focus on variation that would be likely to be seen within MR images.}
	\label{fig:augmentation}
\end{figure}

Convolutional neural networks, however, still ultimately require a reasonable amount of data (100s or 1000s) in the target data domain, as at least some of the network parameters must be fine-tuned to optimise the network for the specific dataset and task. Even though the amount of data required is likely to be reduced (the degree to which it is reduced will be determined by the similarity between the proxy task and target task \citep{He2019, Kornblith_2019_CVPR}) the amount still required may remain greater than is available for the task we are exploring. In this circumstance, data augmentation is normally applied \citep{Simard1998, Chapelle2001}, artificially increasing the size and diversity of the training dataset through applying transformations, creating slightly perturbed versions of the data. 

These augmentations can take the form of basic transformations such as flips and rotations \citep{Krizhevsky2012,Simonyan15} as standardly applied in computer vision tasks, to more extreme examples such as mixup \citep{Mixup} which merges images from different classes together to form hybrid classes, or generative networks such as conditional Generative Adversarial Networks (GANs), which are networks trained to generate simulated data \citep{ConidtionalGAN}. Some example augmentations can be seen in Fig. \ref{fig:augmentation}. While the vast majority of deep learning studies apply data augmentation during training, some studies explore this for neuroimaging specifically, and its effect on model performance. For instance augmentation can be achieved through GANs being used to generate additional meaningful datapoints \citep{WuCGAN,Chaitanya2019, Zhuangaugmentation} or registration to templates \citep{Nguyen2020,Lyu2021}, which generate biologically plausible transformations of the data. Similarly, they can be produced by identifying augmentations which are plausible across sites and scanners \citep{Torchio, Billiot2020}, such as applying bias field (scaling intensities by a smoothly-varying random gain). 

Existing literature suggests that performing augmentations, even transformations which create images beyond realistic variation \citep{Billiot2020}, helps the network to generalise better to unseen data at test time. Data augmentation must be handled with caution, however, so that the transformations applied do not change the validity of the label associated with the image. Consider, for instance, when trying to classify Alzheimer’s disease from structural MRI: the key indicator could be the atrophy of the hippocampus and so, if any transformations are applied during the augmentation process that affect this region (e.g. local elastic deformations), care must be taken to ensure that the perceived level of atrophy is not affected and, thus,  the true label changed. Ensuring this requires high levels of specific domain knowledge and therefore, for certain scenarios, limits the augmentations which can be applied. 

Other approaches to solving the shortage of available training data focus on breaking the input data down into patches e.g. \citep{DeepNat,Lee2020patches} or slices (where 3D data is available) with many studies treating MRI data as 2D inputs e.g. \citep{Livne2019, Fastsurfer2020,Sinha2020}. This approach can vastly increase the amount of available data and can be especially effective for segmentation tasks where we have voxelwise labels. However, by fragmenting the image, these approaches lead to the loss of global information. Although this can be compensated for to some degree using optical flow \citep{Zitnick2005} or conditional random fields \citep{DeepMedic}, patch-wise or slice-wise approaches cannot necessarily be applied to classification tasks, where a single label is provided for the whole image (which may not be valid for all slices) on a given patch or slice of the image \citep{Bijen2018}. When these approaches can be applied, say for a segmentation task, care must be taken when combining the patches at the output to reconstruct a single output image, so that we do not suffer from artefacts at boundaries. Fully 3D networks have been found in most cases to provide better results when they can be implemented \citep{DeepMedic}.

\subsection{Differences between datasets or data domain shift}

Having sufficient data to train the model, however, is only the first difficulty being faced by clinical application of these methods. Deep learning methods have a very high degree of freedom which enables them to learn very complicated and highly non-linear mappings between the input images and the labels,  but this same high degree of flexibility comes at a cost: deep learning methods are prone to \textit{overfitting} to the training data \citep{Srivastava2014}. Furthermore, while, a well-trained model should interpolate well to data which falls within the same distribution as that seen during training, the performance degrades quickly once it has to extrapolate to out-of-distribution data, and even perturbations which are not noticeable by the human eye can cause the network performance to collapse \citep{Papernot2017}. Considerable research effort within computer vision has focused on generalisability from the dataset seen during training to the dataset only seen during testing, where both datasets are drawn from the same distribution. For clinical translatability, we would need generalisability from the training set to all  other reasonable datasets, including those which have not yet been collected. 

\begin{figure}
        \centering
	\includegraphics[width=\textwidth]{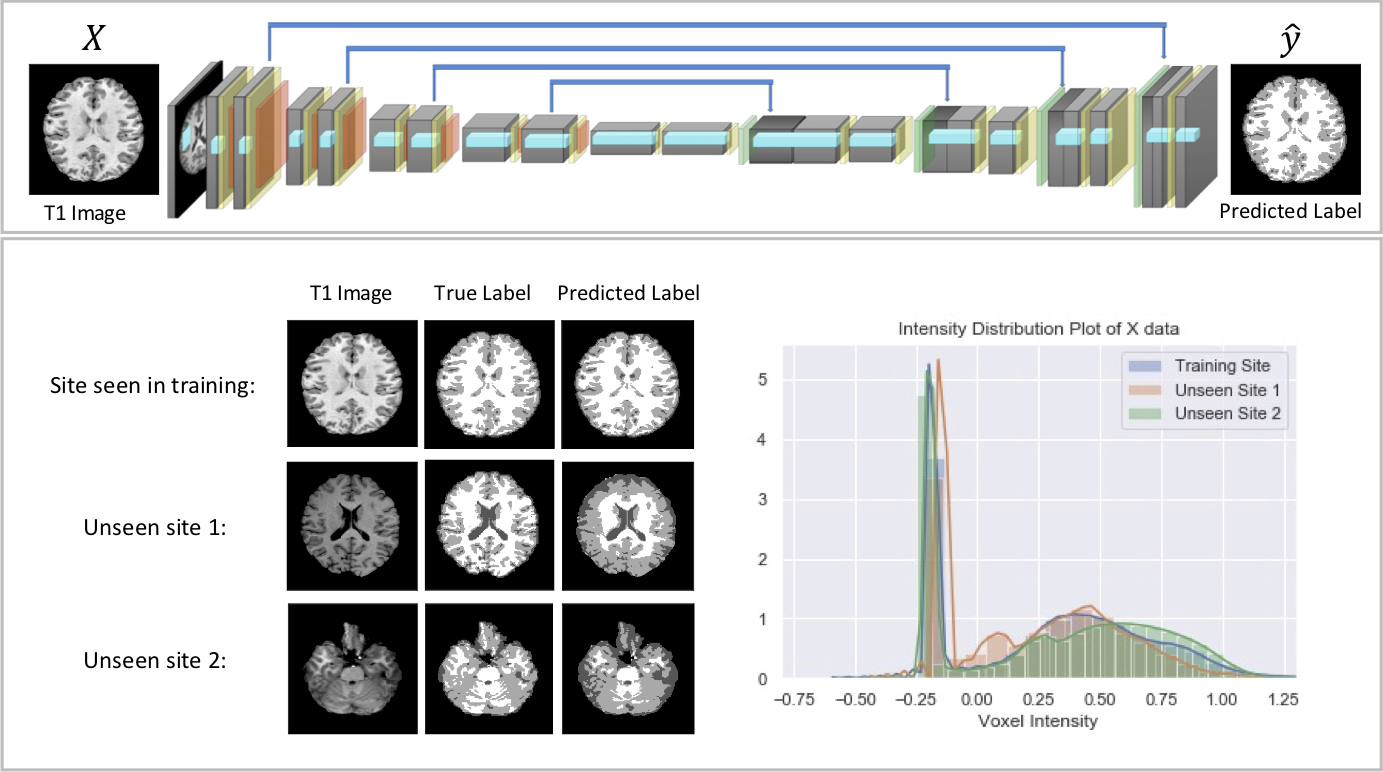}
	\caption{To demonstrate the effect of the difference between domain datasets or domain shift, we completed tissue segmentation on data from three sites collected as part of the ABIDE \citep{ABIDE} multisite dataset. Although the data was collected as part of the same study, there are differences between the data collected at different sites due to being collected on different scanners. The architecture used was a 3D UNet \citep{Cicek2016} with T1 as the input image, and only images from one site were used during training. The predictions can be seen for example images for three sites, the one seen during training and two unseen. It is clear that the segmentation for the site seen during training is good but suffers significant degradation when applied to the unseen sites, despite them being collected for the same study and having similar (normalised) voxel intensities, demonstrating the potential difficulties caused by domain shift. }
	\label{fig:domainshift}
\end{figure}

If, for instance, we consider multisite datasets, such as from the ABIDE study \citep{ABIDE}, where attempts have been made to harmonise acquisition protocols and to use identical phantoms across imaging sites, there is still an increase in non-biological variance when we pool the data across the sites and scanners \citep{Yu2018}. A demonstration of this variance leading to performance degradation for a segmentation task is shown in Fig. \ref{fig:domainshift}.  Multiple studies have confirmed this variation, identifying causes (batch effects) from scanner and acquisition differences, including scanner manufacturer \citep{Han2006, Takao2013}, scanner upgrade \citep{Han2006}, scanner drift \citep{Takao2011},
scanner strength \citep{Han2006}, and gradient non-linearities \citep{Jovicich2006}. 

The removal of this scanner-induced variance is therefore vital for many neuroimaging studies. The majority of deep learning approaches try either  to produce harmonised images \citep{Karayumak2019, stjean2019, Dewey2019, Zhao2019}, or to remove the scanner-related information from the features used to produce the predictions \citep{Moyer2020, Dinsdale2021b}. Both approaches aim for any results obtained for use downstream to be invariant to the acquisition scanner and protocol. These methods succeed in removing the scanner effects from the predictions, but hold no guarantees for scanners not seen during training, and, as the results are very hard to verify without `travelling heads datasets' (images from the same subjects acquired on the scanners to be harmonised) the results obtained from the generated harmonised images are hard to validate \citep{Moyer2020}.

The domain shift experienced when we have multisite data is much less than might be expected when we move between research data and clinical data, or even just two datasets collected independently. The domain shift here can come from two clear sources: the scanner and acquisition, and the demographics (or pathologies) of the studies. Firstly, MRI scans collected for research are often at a higher resolution and higher field-strength than clinical scans. On the other hand, clinical scans are designed to be much more time efficient -- both in terms of acquisition time and in time required for visual inspection -- and are often collected at a lower resolution, often at a lower field strength.   Research scans frequently also have isotropic voxel sizes, whereas anisotropic voxels are still the norm in the clinic and are present in the vast majority of legacy data \citep{Iglesias2020}. Unfortunately, due to the lack of training data already discussed, we are unlikely to be able to train sophisticated models directly on clinical data. 

Therefore, methods are being developed that consider this domain shift  (e.g. between clinical and research data), which focus either on domain adaptation approaches to create shared feature representations for the different datasets, or in synthesising data to enable us to use  the clinical domain. Unlike the harmonisation paradigm, such approaches typically do not wish to allow us to combine the datasets, but simply to be able to harness the shared information from one domain for use in the other. Domain adaptation techniques normally consider the situation where we have a large source dataset -- say a research dataset such as UK Biobank \citep{Biobank} -- and a much smaller target dataset -- say the clinical dataset that we are actually interested in \citep{Ben-David2010, Ganin2015}. Domain adaptation then asks the question: \textit{can a shared embedding be found which is discriminative for the task of interest, while invariant to the domain of the data}? This can take the form of a fully supervised problem, where task labels are available for both datasets, semi-supervised \citep{Valverde2019}, where only a small number of labels are available for the target dataset, or unsupervised \citep{Kamnitsas2017, Perone2019}, where no labels are available for the target dataset. The vast majority of approaches perform domain adaptation across domains only; however, some also consider adaptation across related tasks \citep{Tzeng2015} and have been applied for segmentation \citep{Kamnitsas2017, Valverde2019, Perone2019, Sundaresan2021} and classification problems \citep{Guan2020}. These methods are clearly closely related to transfer learning from a large dataset, but a single feature representation is found for both source and target dataset, rather than creating a new one for the target dataset. These methods perform well on the target data, but further work is required to enable them to adapt reliably to higher numbers of datasets simultaneously. 

Domain adaptation methods, at the extreme, essentially have the end goal that the network would work regardless of the acquisition, which is an active area of research \citep{Billiot2020b, Siddhesh2020, Billiot2020c}. The other approach which has been explored is to use generative methods to convert the data from one domain to the other \citep{Igelsias2020}, such that the transformed data can be used in the existing model. Any generated images must be carefully validated to ensure that they convey the same information as the original and that the outcomes are the same. 

Finally, research data is generally cleaner than clinical data. For instance, many developed algorithms require multiple input modalities and therefore, cannot be applied to a subject if they do not have all the scanning modalities available \citep{Chen2018}. Missing modalities, different fields of view, and incidental findings would all potentially lead to the performance of the model being significantly degraded or the model simply not being applicable. Approaches to deal with missing modalities exist \citep{Zhoumissingmodalities}, but generally still result in a significant degradation in performance, compared to when all modalities are present. Similarly, models are likely to suffer performance degradation or exhibit unexpected behaviour when presented with unexpected pathologies that were not present in the training set \citep{QuickNAT}.

\subsection{Data Composition and Algorithmic Biases}

Finally, and potentially most concerning, is the consideration that the demographics of study data frequently do not fully represent the population as a whole, and so a domain shift is experienced when we attempt to move from the research domain to the clinical domain. Research data is usually acquired with a certain study question in mind and therefore subjects are selected so as  to try to allow targeted exploration of that question. Therefore, research datasets rarely contain subjects with co-morbidities, and subjects with incidental findings would possibly be excluded from the study. For example, patients with advanced Alzheimer's disease are unlikely to be recruited for a general imaging study, due to the ethical implications \citep{Clement2019}, such as the inability to consent and the potential trauma of being scanned. 

In  addition, there exists a strong selection bias, both in relation to the people who volunteer for studies and those who see the study through to the end, with studies having demonstrated associations to age, education, ancestry, geographic location, and health status \citep{Karlawish2003, Clement2019}.  Furthermore, people with family connections to a given condition are more likely to volunteer for a study as a healthy control, leading to certain genetic markers being more prevalent in a study dataset than in the population as a whole \citep{Hostage2013}. Therefore, associations learned when considering research data may not generalise, and care must be taken in extrapolating any model trained on these datasets to clinical populations. 

Therefore, any trained model will suffer from algorithmic bias: that is, the outcomes of the model will potentially be systematically less favourable to, or have systematically lower performance on, individuals within a particular group, where there is no relevant difference between groups that justifies such effects \citep{Paulus2020}. Erroneous or unsuitable outcomes will likely be produced for the groups who are less likely to be represented in the training data. Algorithmic bias is therefore a function of the creator and the creation process, and fundamentally, of the data which drove the model training. When considering bias, there are two issues that must be considered: first, \textit{does the algorithm have different accuracy rates for different demographic groups}? Second, \textit{does the algorithm make vastly different decisions when applied to different populations}? As networks simply learn the patterns in the data, any bias, such as racial bias \citep{Williams2015}, in the data may be learnt and encoded into the models. 

Inevitably, when considering complicated questions with extremely heterogeneous populations, the datasets used to train the deep learning methods will be incomplete and insufficient in terms of spanning all the possible modes of variability \citep{Ning2020}. For instance, pathologies will occur against a background of normal ageing, with differences being present between individuals due to both processes. Sufficiently encompassing all this variation is infeasible, not only due to the number of subjects which would be required, but also to the difficulty in recruiting subjects from some specific groups. There is, therefore, often inadequate data from minorities, and consequently any model found through the optimisation process (which is usually selected as the model which performs best on the average subject in the validation dataset) will probably be inadequate for these groups, especially for conditions with higher rates in these groups \citep{Wong2015}. When models are developed for clinical translation, therefore, the limitations of the models must be understood: wherever groups are under-represented,  the appropriateness of the application of the model must be considered, and any limitations identified. Where these limitations mean that minorities will receive a lower standard of care, the models are inappropriate.

\section{Interpretability}

\begin{figure}[h]
        \centering
	\includegraphics[width=0.7\textwidth]{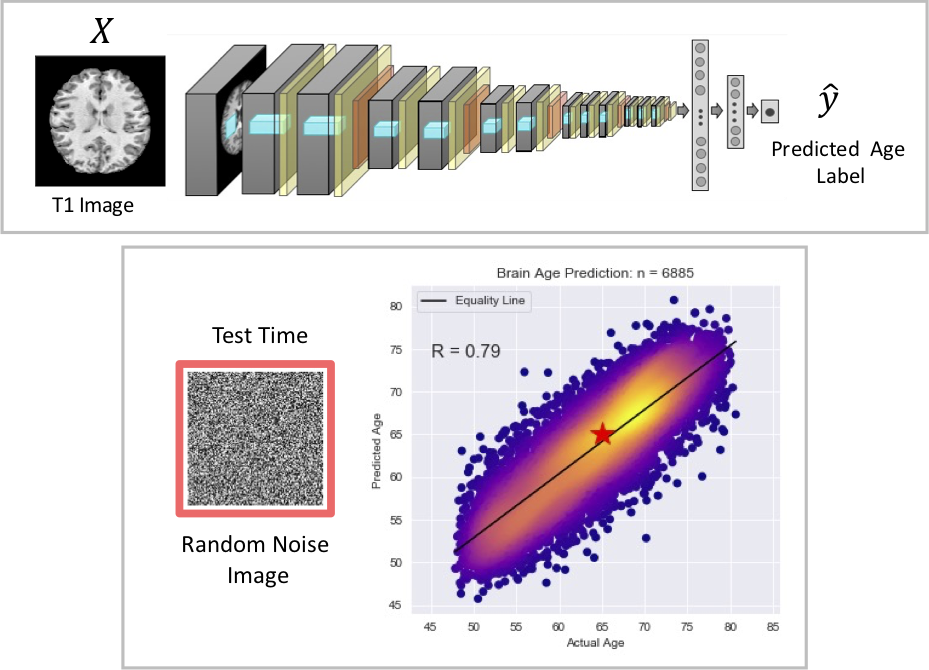}
	\caption{If we take a model trained to predict brain age \citep{Dinsdale2021} from T1 structural images and present the network with an image of random noise, as it is unable to output an unknown class, the network predicted the random noise to have an age of 65, around the average age of the subjects of the dataset. While we clearly can identify the random noise image, there are many situations where the model could be presented with an image outside of the distribution it was trained on, but would still output a (meaningless) result, which would be much harder for a user to identify.}
	\label{fig:randomnoise}
\end{figure}

The performance degradation experienced with domain shift would be potentially less problematic were it not for another problem associated with deep learning methods: that is, models will output a prediction for any data fed in, but that prediction may not necessarily be meaningful. Lacking a `do not know’ option, given any image of the correct input dimension, a neural network will output a prediction, even if that prediction is meaningless or the input is nonsensical. For instance, if a random noise image is fed into a network trained to predict brain age, the network will predict an apparently valid age for the random noise (see Fig. \ref{fig:randomnoise}). While in this example, visually identifying the pure-noise image is trivial, were the network trained for a more complicated classification task, identifying erroneous results is more difficult and requires large amounts of clinical and domain knowledge. This therefore leads to the critical question, can the results be trusted? 

Despite  the assertion of Geoffery Hinton, head of GoogleBrain, that there is no need for AI to be interpretable\footnote{\url{https://www.wired.com/story/googles-ai-guru-computers-think-more-like-brains/}}, the majority would agree that, were deep learning and AI methods to be used to determine patient care, they need to be interpretable and interrogable. Interpretability is often defined as \textit{`the ability to provide explanations in understandable terms to a human’}. The explanations should, therefore, be logical decision rules which lead to a given diagnosis or patient care being chosen, and the understandable terms need to be from the domain knowledge related to the task. This is especially important because neural networks have no semantic understanding of the problem. That is, they have no understanding of the problem they are being asked to solve. Rather, they are blunt tools which, given $\bm{X}$ and $\bm{y}$, learn a mapping between the two. If there exists spurious information in $\bm{X}$ which can aid in this mapping (or \textit{confounders}), then this information will probably be used, misleading the predictive potential of the network. Consider for instance, the case where all subjects with a given pathology were collected on the same scanner. A network could then achieve $100\%$ recall (correctly identify all examples) for this pathology by fitting to the scanner signal, rather than learning any information about the pathology \citep{Winkler2019}. It would then, in all probability, identify a healthy control from the same scanner as having the same pathology. 

The effect of confounders would not be observed without further probing of the behaviour of the trained model, and the probing of networks is non-trivial. This has led to neural networks being commonly described as ‘blackbox’ methods. There is therefore a need for interpretable networks, allowing both understanding and scrutiny of any decision made. Approaches have broadly focused on two main areas: visualisation and uncertainty.  

\subsection{Visualisation}

\begin{figure}
        \centering
	\includegraphics[width=0.7\textwidth]{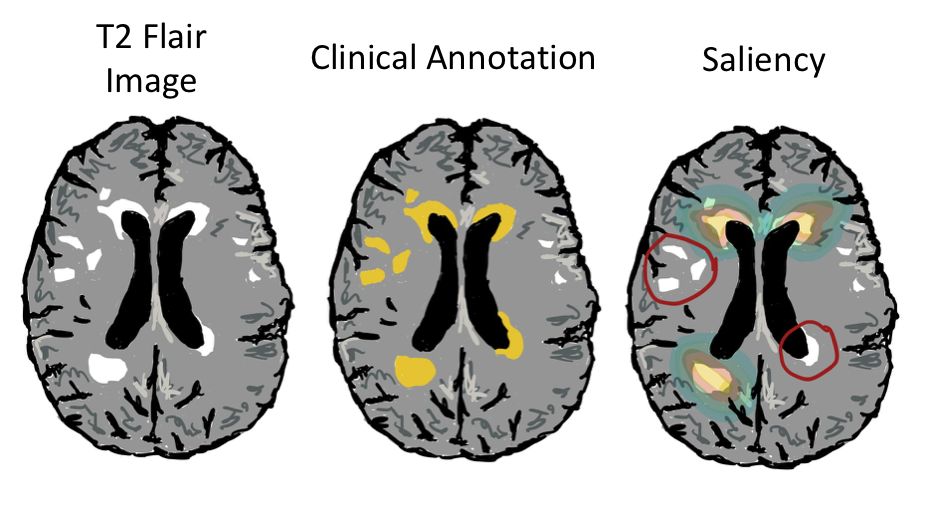}
	\caption{Schematic of the limitation of using saliency: when identifying the presence of white matter hyper-intensities, the neural network might only need to focus on a few of them to make the prediction. This would lead to not all of the white matter hyper-intensities being indicated in the saliency map and so the prediction not matching the clinician's expectation.}
	\label{fig:saliency}
\end{figure}

Visualisation methods generally attempt to show which aspects of the input image led to the given classification, often by creating a `heat map’ of importance within the input image. Many of these methods are post-hoc, taking a pre-trained model and trying to understand how the decision was made. Most commonly, these methods analyse the gradients or activations of the network for a given input image, such as saliency maps \citep{Simonyan14a}, GradCAM \citep{Ramprasaath2017}, integrated gradients \citep{Sundararajan2017}, and layerwise relevance propagation \citep{Binder2016}. They have been applied in a range of MRI analysis tasks to explain decision-making, such as in Alzheimer's disease classification \citep{Bohle2019}, brain age prediction \citep{Dinsdale2021}, and brain tumour detection and segmentation \citep{Mitra2017, Takacs2018}. Other methods are occlusion- (or perturbation-) based, where parts of the image are removed or altered in the input image, then heat maps are generated which evaluate the effect of this perturbation on the network’s performance \citep{Zeiler2014, Abrol2020}. Most of these methods, however, provide coarse and low resolution attribution maps and can be extremely computationally expensive \citep{Bass2020}, especially when working with 3D medical images.

These posthoc methods do not require any model training in addition to the original network, however, they have been shown to be, in some instances, unable to identify all of the salient regions of a class, especially in medical imaging applications \citep{Bass2020,Baumgartner2018}. It has been shown that classifiers base their results on certain salient regions, rather than the object as a whole, and therefore, a classifier may ignore a region if the information there is redundant -- i.e. if it can be provided by a different region of the image, which is sufficient to minimise the loss function. Therefore, the regions of interest highlighted by these methods may not fully match the expectations of a clinician (see Fig. \ref{fig:saliency}), and also the prediction results might be virtually unchanged if the network was retrained with supposedly-salient areas removed. Generally, although many methods have been developed to produce saliency or `heatmaps' from CNNs, limited effort has been focused on their evaluation with actual end-users \citep{Alqaraawi2020}. Furthermore, these methods, at their best, only highlight the important content of the image, rather than uncovering the internal mechanisms of the model. 

Attention gates are a component of the network, which aim to focus a CNN on the target region of the image  (the salient regions) by suppressing irrelevant feature responses in feature maps during training rather than post-hoc. This focuses the attention of the network onto the information critical for the specific task, rather than learning non-useful information in the background or surrounding areas \citep{Park2018,Wang2018,Hu2018}. These methods provide the user with attention maps, which again highlight the regions of the input image driving the network predictions. However, these methods, similarly to saliency or gradient based methods, may not highlight all of the expected regions in the image. Attention gates have been applied to a range of medical image analysis tasks, both for classification \citep{Dinsdale2021,Schlemper2019} and segmentation \citep{Schlemper2019, Zhuang2019, Zhang2020}. Other methods have been developed to allow the visualisation of the differences between classes directly, rather than analysing the model post-hoc \citep{Bass2020,Lee2020}. Other methods aim to increase their interpretability by breaking down the task into smaller, more understandable tasks, such as first segmenting a region known to be a biomarker for a given condition, and then classifying based on this region \citep{Lui2020}.

The  methods discussed so far enable the visualisation of the regions of the input image  which drive the predictions, but they do not provide insight into how the underlying filters of the network create decision boundaries. In addition, in brain imaging, the class phenotypes are typically heterogenous and any changes they cause probably occur simultaneously, with significant amounts of healthy and normal variation in shape and appearance, meaning that interpretation of feature attribution maps to understand  network predictions is often difficult. Given the millions of parameters in many deep learning networks, despite our ability to visualise individual filters and weights helping us to understand the hierarchical image composition,  it is very difficult to interrogate how decision boundaries are formed. Therefore, there is  a need to create networks which complete their predictions in more understandable ways, without restricting the complex non-linearities in behaviour that are necessary for good prediction performance. 

\subsection{Uncertainty}

\begin{figure}[h]
        \centering
	\includegraphics[width=0.9\textwidth]{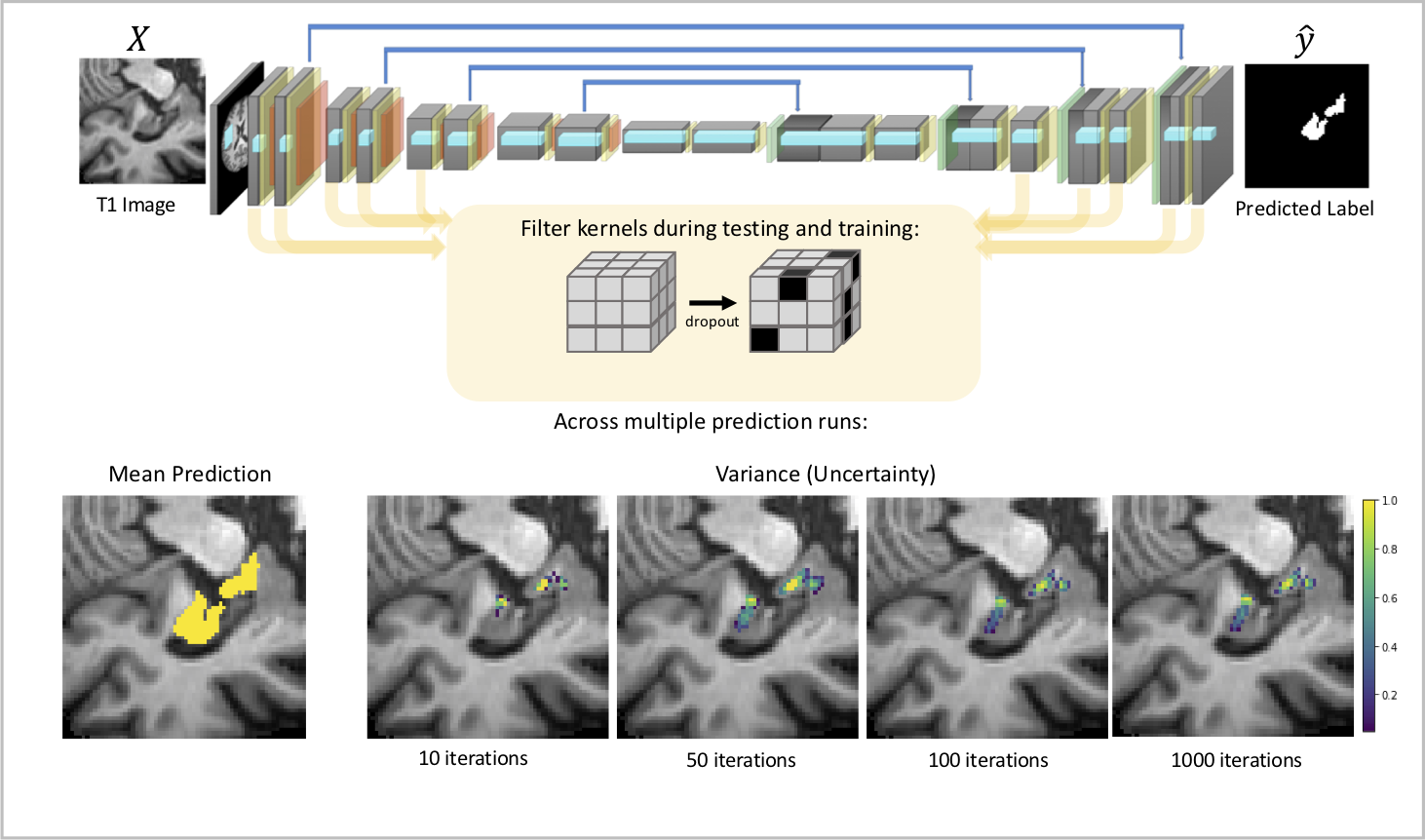}
	\caption{Methods which use Monte Carlo dropout have dropout applied at training and test time, where weights in the convolutional kernels are removed, which is approximated to represent the distribution of possible model architectures at test time. To demonstrate this we trained a standard 3D UNet to complete hippocampal segmentation, with a dropout value of 0.5 applied on all convolutional layers. The HarP dataset \citep{HARP} was used in this experiment, preprocessed as in \citep{Dinsdale2019}. For each subject, we obtained a mean prediction and a uncertainty map, indicating the regions where the predictions between models were the most varied and so approximated to be the least certain.}
	\label{fig:uncertainty}
\end{figure}

The use of uncertainties, then, is an approach which aims to address the problem that, regardless of the input image, neural networks will always output a prediction, even when they are very unsure of the prediction. An example is when the data falls far away from the domain of data the network was trained on. Furthermore, the softmax values frequently output by a neural network are not true probabilities \citep{Gal2016}, and networks often output high, incorrect softmax values, especially when presented with noisy or ambiguous data, or when the data presented to them differs from the distribution of the training data. Therefore, the development of models aware of the uncertainty in their predictions is key for providing confidence and trust in systems -- and this is not provided by traditional deep learning algorithms. 

Uncertainties in deep learning can be split into two distinct groups \citep{Kendall2017}: \textit{aleatoric uncertainty}, the uncertainty due to the ambiguity and noise in the data, and \textit{epistemic uncertainty}, which is due to the model parameters. The majority of methods in the literature focus on epistemic uncertainty, using Bayesian approaches to quantify the degree of uncertainty. The goal here is to estimate the posterior distribution of the model parameters. However, due to the very high dimensional parameter space, analytically computing the posterior directly is infeasible. Therefore, most methods use Monte Carlo dropout \citep{Gal2016}, demonstrated in Fig. \ref{fig:uncertainty}, where dropout \citep{Srivastava2014} is applied to each of the convolutional layers and kept at test time, and thus the model architecture is random at test time. Although other methods exist, in the majority of approaches for medical imaging, the uncertainty is then quantified through the variance of the predictive distribution, resulting from multiple iterations of the prediction stage with dropout present at test time, as demonstrated in Fig. \ref{fig:uncertainty}. This approach can be simply applied to existing convolutional neural networks and in medical imaging has primarily been used for segmentation tasks \citep{Roy2018, Kwon2020}, where the segmentation is predicted alongside an uncertainty map. Other works have studied disease prediction, where the uncertainty is associated with the predicted class \citep{Jungo2018SurvivalPred,Tousignant2019} and image registration \citep{Bian2020}. However, care must be taken with choice of the hyperparameters (the dropout probability and the number of iterations that the variation is calculated over) to ensure that the model assumptions are reasonable.

Some methods focus on the aleatoric uncertainty instead, which is estimated by having augmentation at test time \citep{Ayhan2018,Wang2019}. Understanding of the uncertainty introduced by data varying from the training distribution is vital for clinical translation of deep learning techniques: with the degree of variation present in clinical data between sites and scanners, it is vital to understand what this variation adds to predictions, both to allow it to be mitigated against, and to provide users with confidence in the predictions. Unsurprisingly, there is a correlation between erroneous predictions and high uncertainties, and so, this could be used to improve the eventual predictions \citep{Jungo2018SurvivalPred,Herzog2020}. 

There is, however, need for further development of these methods to ensure that the uncertainties produced are meaningful in all the circumstances in which they could be deployed. For instance, further study is required to ensure that the uncertainty is meaningful for all possible dataset shifts and to provide a calibration for the uncertainty values so that they are comparable across methods \citep{Thagaard2020, laves2020wellcalibrated}. Furthermore, the uncertainty values are, also, only as good as the trained model, the assumptions behind the uncertainty model, and are only meaningful alongside a well-validated model which is sufficiently powerful to discriminate the class of interest.

\subsection{Interrogating the Decision Boundary}
For many applications in medical imaging, the output of the deep learning algorithm, if applied clinically, could potentially directly influence patient care and outcomes. Therefore,  there is a clear need to be able to interrogate how decisions were made \citep{Shah2019}. For instance, if we reconsider Fig. \ref{fig:decisionboundary}, it is clear that the location of the decision boundary could impact highly on the care for the patient if the classification was the presence or absence of white matter hyper intensities. While visualisation methods allow inspection of which regions of the image influenced the prediction, and uncertainties grant us an insight to the confidence we should place in a given prediction, for many applications we need to know precisely which characteristics led to a given prediction. This is also important to help with the identification of algorithmic bias influencing the decision making. 

One method of understanding the decision boundary is \textit{counterfactual analysis}, which, given a supervised model where the desired prediction has not been achieved, shows what would have happened if the input to the model were altered slightly \citep{Verma2020}. In other words, it identifies what altered characteristics would have led to a different model prediction. However, applications to neuroimaging \citep{Pawlowski2020}, and even medical imaging \citep{Major2020, Singla2021} more generally, are currently few and the utility across neuroimaging tasks needs to be explored.

\section{Evaluation}
\subsection{Availability of Training Labels}
The evaluation of metrics also requires labels -- that is, a \textit{ground truth} label created by a domain expert. In medical imaging, the ground truth is regarded as labels created by domain experts and these labels are key for training models, but do not necessarily form part of standard clinical practice. Firstly, the labels are required for evaluation of the model performance, and secondly, they are required to allow the training of supervised methods. This, therefore, exacerbates the problem of the shortage of data as we need not only large amounts of data, but we also need equal amounts of correct manual labels. These labels are expensive to obtain, requiring large allocation of expert time to curate and expert domain knowledge. Therefore, there is a need to develop methods which can work in data domains where low numbers of labelled data points are available. 

Few- and zero-shot learning methods work in very low data regimes and are beginning to be applied to medical imaging problems \citep{feyjie2020}. They are, however, very unlikely to generalise well to images from other sites and scanners, as the variation seen will not span the expected variation of the data but they can help begin to learn clusters within subjects where few labels are available. Unsupervised domain adaptation has been applied more widely, including for neuroimaging problems, to help cope with a lack of labels, with information from one dataset being leveraged to help us perform the same or a related task on another dataset \citep{Perone2019, Sundaresan2021}.  

Other methods to overcome the lack of available labels focus on working with approximations for labels, which are cheaper to acquire \citep{Tajbakhsh2020}. For instance, many methods propose pre-training the network using auxiliary labels generated using automatic tools (e.g., traditional image segmentation methods) and then fine-tuning the model on the small number of manual labels \citep{QuickNAT, Wang2020}, or registration of an atlas to propagate labels from the atlas to the subject space \citep{Zhu2019atlas}. Other approaches are weakly supervised, utilising quick annotations such as image level labels \citep{Feng2017}, bounding box annotations \citep{Rajchl2017},  scribbles \citep{Dorent2020,Luo2021} or  point labels \citep{Mcever2020}.

Other approaches to allow us to utilise deep learning when we have access to limited numbers of training labels include active learning and omnisupervised learning, both trying to make the most effective use of the limited number of labels available for a task. Active learning aims to minimise the quantity of labelled data required to train the network by prompting a human labeller only to produce additional manual labels where they might provide the greatest performance improvements. This minimises the total number of annotations that need to be provided, and provides better performance than randomly annotating the same number of samples \citep{Lin2017, Nath2021}. In omnisupervised learning \citep{Radosavovic2018} automatically generated labels are created to improve predictions, starting from a small labelled training set. By combining data diversity through applying data augmentation, and model diversity through the use of multiple different models, a consensus of labels is produced, which can be used to train the final model. For both approaches, the labels used are chosen using various different approaches such as uncertainty \citep{pmlr-v70-gal17a, Venturini2020}.
\\

The difficulty in acquiring good quality manual labels, is of course, exacerbated by the variance caused when we pool data, as discussed above. This increase in variance limits the impact any produced labels can have, and so again, methods of pooling data across datasets without getting an increase in variance due to the scanner effects will be necessary. 

The labels themselves, however, will provide another source of variance \citep{Cabitza2019}: when working with medical images, the labels are frequently complicated and ambiguous \citep{shwartzman2020worrisome}, often open to interpretation or with subjects having multiple labels that could be attributed due to co-morbidities \citep{Graberii21} but despite this we usually assume them to be $100\%$ accurate \citep{Cabitza2020} -- the `gold' standard. If there is no objective answer, we cannot expect networks to provide an objective answer. Furthermore, this also leads to inter-\textit{rater} variability between the people creating the labels, which leads to a degree of uncertainty in the produced ground truth. The effect that this variability has on the predictions of the network needs to be understood and mitigated against, with the challenge being that obtaining labels for multiple raters for datasets is deeply infeasible, and has been shown to limit the performance of algorithms \citep{svensson2015}. The uncertainty in the labels is also amplified by the lack of available data for rare conditions, which are therefore less represented in datasets, resulting in raters having less experience assessing them. This is especially problematic if trying to quantify longitudinal changes with different raters \citep{Visser2019}. 

Approaches need to consider three factors \citep{Cabitza2020}: \textit{agreement} -- the degree to which raters agree on a given label; \textit{confidence} -- how certain a rater is in their label, and \textit{competence} --  how accurate a given rater is. Research directions into the effect of rater variance have largely focused on either quantifying the reliability of the labels \citep{Cabitza2020}, or quantifying its effect on network performance \citep{shwartzman2020worrisome, Visser2019, Haarburger2020}. Before any algorithm is deployed in practice, the limitations due to the labels must be understood, and its consideration become a standard part of any deployment pipeline, remembering always the \textit{``garbage in, garbage out"} principle, for instance by producing the range of segmentations that might be produced when training with multiple raters \citep{Baumgartner2019}.

\subsection{Choice of Loss Function}
When training networks and evaluating the model performance, we have to choose a loss, or cost, function which the network aims to minimise. Although some works design bespoke, task-specific cost functions, the majority are based on standard functions, such as categorical cross entropy for classification and segmentation tasks, Dice (a measure of overlap between the  predicted and true label) for segmentation, and mean square error (MSE) for regression-based tasks. 

These metrics are normally chosen because of their well-understood and characterised behaviour \citep{MaierHein2018}, and, especially in the cases of segmentation, their ability to produce visibly acceptable predictions. For clinical translation of deep learning methods, however, we rather need to consider which measures are most important for the clinical application \citep{Shah2019, Keane2018}. Metrics only tell us part of the story and it is important that we ensure that all vital information for clinical assessment is provided by the reported metrics. For instance, in many cases, false negatives are more problematic than false positives, resulting in a patient failing to receive the necessary care. Similarly, when considering tasks such as tumour segmentation, the classification of the center of the tumour is trivial and the utility for a clinician would be identification of boundaries, where the classification is harder or ambiguous. Therefore, a metric such as the Dice overlap metric is not necessarily a good indicator of performance from a clinician's point of view, with the boundary having little impact on the metric if the region of interest is large, while being potentially critical for patient care. Developing networks and loss functions with the specific application in mind is vital and has been explored for some applications, such as focusing segmentations to the boundary \citep{QuickNAT, Hatamizadeh2019, Zhang2020boundary}, but needs to be considered for each application in turn.

Furthermore, when training neural networks, we generally maximise the performance averaged over the batch (a subset of the training data). In practice, however, we are more likely to care about the performance on the hardest examples being acceptable, than the performance on the easiest set of examples being improved slightly  \citep{Shu2020}. If we, for instance, consider a segmentation task, a given level of performance could be achieved by easier examples being perfectly segmented, and more difficult examples being completely missed. We would, however, probably prefer trading-off a small amount of performance on the easier examples in return for better performance on the harder examples, which may give the same average performance overall.  Therefore, the standard practice of minimising the average performance may not be appropriate when the aim is clinical deployment, rather than the development of methods. 

\section{Logistical Challenges}
\subsection{Computational Power}
The final category of challenges are far more logistical. Firstly, most deep learning models are developed using GPUs (graphical processing units), hardware that is rarely available in a clinical setting, and producing predictions on standard hardware often takes infeasible amounts of time or requires prohibitive amounts of computational power. Many of the most successful methods applied in imaging challenges involve large ensemble models such as \citep{Meenakshi2019, Isensee2021}, leading to a very large amount of parameters and therefore calculations that must be stored and computed. 

Therefore, for clinical translatability, methods need to be developed with consideration of the computational limitations that will be present on deployment, and seeking to create solutions which work within these constraints.  Student-teacher networks \citep{Hinton2015, Ghafoorian2018} and model distillation \citep{pmlr-v121-murugesan20a, Vaze2020} aim to create smaller networks which are able to mimic the performance of the original large model, thus reducing the number of parameters in the final network which is deployed. Other approaches use separable convolutions which drastically reduce the number of parameters in the network. Model pruning \citep{OptimalBrainDamage, Molchanov2016, Li2017} acknowledges that the parameters in neural networks are very sparse and, therefore, by removing those that are contributing the least to the final prediction, we can reduce the size of the model architecture whilst having limited impact on the final performance of the model. 

\subsection{Data Sharing and Data Privacy}
\begin{figure}[h]
        \centering
	\includegraphics[width=0.7\textwidth]{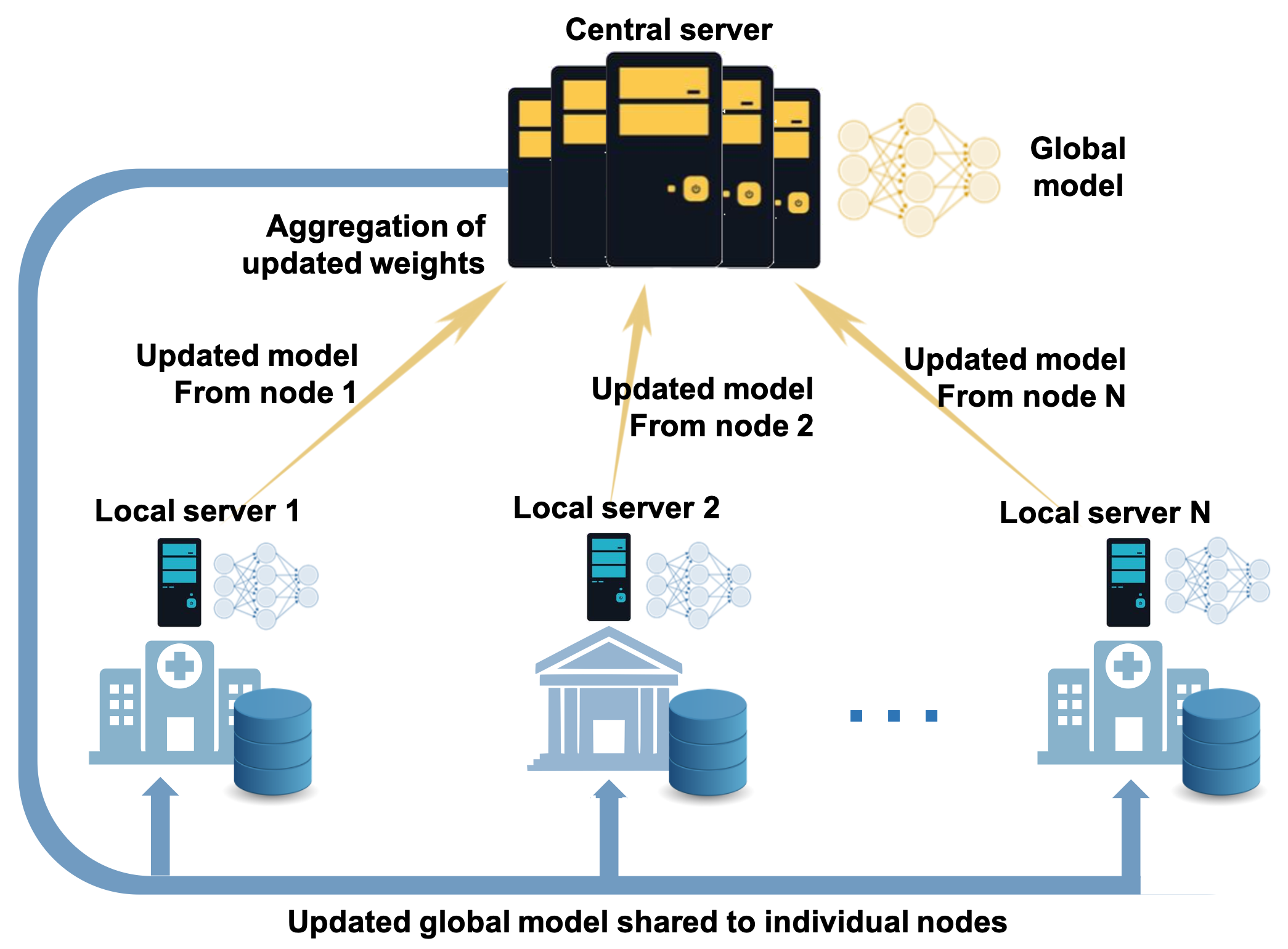}
	\caption{Illustration of a centralised federated learning framework. In the framework, the data for training the model is stored in local servers and not shared with the central server to ensure data security. On the other hand, while the global model is available in the central server, the model parameters are shared with the local nodes 1,2... N (indicated by blue arrows) where training and parameter updates happen. The updated weights are then received at the central server (indicated by yellow arrows), where the incoming updates are aggregated and applied to the global model. This learning and update happens in an iterative manner, and both up and down transfer of model parameters are encrypted for data security. }
	\label{fig:federatedlearning}
\end{figure}

Clearly from the above discussion of data and labels, if we want CNNs which work for patients in real clinical applications, we need to be able to train our models on medical data which is relevant, realistic, and representative. Many current approaches focus on pooling anonymised data from across sites and patient groups,  and to be able to do this concentrate on removing identifiable features such as name, birth data and faces from the images. However, neural networks are still capable of extracting identifiable features from these anonymised images such as age and sex \citep{Pawlowski2019}, which, in combination with other features such as hospital location and illness, could be identifying \citep{Sweeney2002}. As neural network research continues to develop, it is probable that their ability to extract information despite the best efforts at anonymisation will increase. Furthermore, a proportion of identification risk comes from the presence of other auxiliary information, for instance, in neuroimaging, the scanner used to acquire the image. This is known as \textit{linkage attack}, and is increasingly difficult to protect against across fields  using classic anonymisation techniques \citep{Sweeney2002,Gentilli2017, Bindschaedler2018}.

While de-identifying this data may just seem like an extra task for medical researchers, there are parties whose core business model is to de-anonymize medical data that has been sold for research and sell that information to insurance companies \citep{Tanner2017}. De-anonymisation research is a rapidly advancing field: for instance, reconstructing the faces of defaced medical images \citep{Abramian2019,Vidanage2020}. Therefore, in order to avoid data privacy problems in the future, approaches avoiding the aggregation of private medical information are valuable. 

Fortunately, medical research is not the only field to face difficulties regarding the handling of sensitive and personal information. For instance, banking and mobile phone companies have faced this problem before: \textit{how can one extract only the desired answer, without also collecting information that could be used for harm}? Therefore, we can take advantage of the privacy-preserving data analysis techniques that have rapidly developed in recent years.  The field of privacy-preserving data analysis consists of a collection of techniques that allow models to be trained without having direct access to the data, and that prevent these models from inadvertently storing  sensitive information about the data. The most popular of these techniques are \textit{federated learning}, \textit{differential privacy}, and various forms of encrypted computation \citep{AlRubaie2019, Kaissis2021}. Here we will focus on federated learning and differential privacy, as they currently show the most practical relevance in a neuroscience research setting \citep{Reike2020}; for a broader review see, for instance, \citep{Kaissis2021}.
\\

Federated learning means training or testing your model (or performing your statistical query) on data that is stored on different devices or servers across the world, without having to centrally collect the data samples into one local aggregate dataset \citep{LiTian2020, Yang2019Fed}.  Instead of moving the data to the model, copies of the global model are sent to where the data is located; the data remains on the hospital server. A model is sent to the device and trained on the local data, after which the newly improved model with its update is sent back to the main server to be aggregated with the main model. This preserves privacy in the sense that the data has not been moved from the device, and therefore is gaining popularity in various healthcare applications \citep{Reike2020} -- for instance \citep{Huang2019,Sheller2019, Wenqi2019}. However, federated learning is limited by the fact that the content of the local data can sometimes be inferred from the weight updates or improvements in the models \citep{Wang2019}. In addition, deep neural networks are often over-parametrised, meaning that they can encode more information than is necessary for the prediction task, resulting in a model that can potentially, inadvertently memorize individual samples. Therefore, to prevent the possibility of inferring personal characteristics from the data, further techniques need to be employed, such as differential privacy \citep{Dwork2014}.

Differential privacy techniques work by injecting a controlled amount of statistical noise to obscure the data contributions from individuals in the dataset \citep{Dwork2014}, and has been applied to medical imaging \citep{Ziller2021}. This is performed while ensuring that the model still gains insight into the overall population, and thus provides predictions that are accurate enough to be useful. Differential privacy is a mathematical framework, and thus enables the degree of privacy loss to be calculated and evaluated based on the concept of a privacy `budget'. Therefore, ultimately the use of differential privacy is a careful trade-off between privacy preservation and model utility \citep{Dwork2014,Shokri2015,Abadi2016}. Importantly, a critical aspect of differential privacy is its inherent robustness to linkage attacks \citep{Sweeney2002}.

\section{Conclusion}
The combination of deep learning based methods and large scale imaging datasets, such as  UK Biobank, offers many opportunities to neuroimaging. Clearly, however, for the full impact of these methods to be experienced in the clinical domain there are challenges that must still be overcome. Ultimately, for models to be able to be deployed successfully, the clinical needs and limitations must be considered centrally to model design, so that the models produced are robust, reliable, and able to improve patient outcomes.

The code for the examples in this paper can be found at \url{github.com/nkdinsdale/challenges_review}. 

\section{Acknowledgements}

This work was supported in parts by  funding from the Engineering and Physical Sciences Research Council (EPSRC) and Medical Research Council (MRC) [grant number EP/L016052/1] (ND and EB), by the Clarendon Scholarship fund (EB), and and a Wellcome Trust Strategic Award (215573/Z/19/Z) (SS). VS is supported by the Wellcome Centre for Integrative Neuroimaging [Grant number 203139/Z/16/Z]. A.N. is grateful for support from the UK Royal Academy of Engineering under the Engineering for Development Research Fellowships scheme. M.J. is supported by the National Institute for Health Research (NIHR) and the Oxford Biomedical Research Centre (BRC). The Wellcome Centre for Integrative Neuroimaging is supported by core funding from the Wellcome Trust (203139/Z/16/Z).

This research has been conducted in part using the UK Biobank Resource under Application Number 8107. We are grateful to UK Biobank for making the data available, and to all UK Biobank study participants, who generously donated their time to make this resource possible. Analysis was carried out on the clusters at the Oxford Biomedical Research Computing (BMRC) facility and FMRIB (part of the Wellcome Centre for Integrative Neuroimaging). BMRC is a joint development between the Wellcome Centre for Human Genetics and the Big Data Institute, supported by Health Data Research UK and the NIHR Oxford Biomedical Research Centre. The views expressed are those of the author(s) and not necessarily those of the NHS, the NIHR or the Department of Health.

The primary support for the ABIDE dataset by Adriana Di Martino was provided by the (NIMH K23MH087770) and the Leon Levy Foundation. Primary support for the work by Michael P. Milham and the INDI team was provided by gifts from Joseph P. Healy and the Stavros Niarchos Foundation to the Child Mind Institute, as well as by an NIMH award to MPM ( NIMH R03MH096321). 

\bibliographystyle{bibfile}
\bibliography{references}

\end{document}